\documentclass[%
 preprint,
groupedaddress,
superscriptaddress,
 amsmath,amssymb,
 aps,
 prl,
 showpacs
]{revtex4-2}
\usepackage{graphicx}% Include figure files
\usepackage{makecell,booktabs}
\usepackage{dcolumn}% Align table columns on decimal point
\usepackage{bm}% bold math
\usepackage{hyperref}% add hypertext capabilities
\hypersetup{
    colorlinks=true,
    urlcolor= blue,
    citecolor=blue,
linkcolor= blue}
\usepackage{xcolor}
\begin{document}
%\begin{CJK*}{GB}{} % Use default fonts from CJK (see below)
%\preprint{APS/123-QED}

\title{Thin film growth of the Weyl semimetal NbAs}% Force line breaks with \\
%\thanks{A footnote to the article title}%

\author{Wilson Yanez}
\author{Yu-Sheng Huang}
 \affiliation{Department of Physics, Pennsylvania State University, University Park, Pennsylvania 16802, USA}%Lines break automatically or can be forced with \\
\author{Supriya Ghosh}
\affiliation{%
Department of Chemical Engineering and Materials Science, University of Minnesota, Minneapolis, Minnesota 55455, USA
}%
\author{Saurav Islam}
\author{Emma Steinebronn}

 \affiliation{Department of Physics, Pennsylvania State University, University Park, Pennsylvania 16802, USA}%Lines break automatically or can be forced with \\

 \author{Anthony Richardella}
 \affiliation{Department of Physics, Pennsylvania State University, University Park, Pennsylvania 16802, USA}%Lines break automatically or can be forced with \\

\author{K. Andre Mkhoyan}
\affiliation{%
Department of Chemical Engineering and Materials Science, University of Minnesota, Minneapolis, Minnesota 55455, USA
}%
\author{Nitin Samarth}
\email{nsamarth@psu.edu}
\affiliation{Department of Physics, Pennsylvania State University, University Park, Pennsylvania 16802, USA}

\date{\today}

\begin{abstract}
We report the synthesis and characterization of thin films of the Weyl semimetal NbAs grown on GaAs (100) and GaAs (111)B substrates. By choosing the appropriate substrate, we can stabilize the growth of NbAs in the (001) and (100) directions. We combine x-ray characterization with high-angle annular dark field scanning transmission electron microscopy to understand both the macroscopic and microscopic structure of the NbAs thin films. We show that these films are textured with domains that are tens of nanometers in size and that, on a macroscopic scale, are mostly aligned to a single crystalline direction. Finally, we describe electrical transport measurements that reveal similar behavior in films grown in both crystalline directions, namely carrier densities of $\sim 10^{21} - 10^{22} $

%We use advanced X-ray characterization to understand the macroscopic nature of our films and show that NbAs (001) grows with a 45$^\circ$ rotation with the basal plane and that there is twinning in NbAs (100). On the microscopic scale, transmission electron microscopy shows the presence of nanometer size domains with a define crystalline orientation that is oriented in different crystalline directions. 
%\begin{description}
%\item[Usage]
%Secondary publications and information retrieval purposes.
%\item[Structure]
% %You may use the \texttt{description} environment to structure your abstract;
% %use the optional argument of the \verb+\item+ command to give the category of each item. 
% %\end{description}
 \end{abstract}
\maketitle

 %\keywords{Suggested keywords}%Use showkeys class 
%\section{\label{sec:level1}Introduction}
% Weyl semimetals 

The TX class (T=Ta/Nb X=As/P) of transition metal monopnictides is a promising quantum materials platform for studying topological phenomena of contemporary interest because the bulk band structure corresponds to that of a topological Weyl semimetal \cite{doi:10.1126/science.aaa9297,PhysRevX.5.031013,PhysRevLett.116.096801,Huang2015}. Broken inversion symmetry in this class of materials leads to Weyl nodes with different chiralities; when projected onto the surface of the crystal, these nodes are connected by topological surface states (Fermi arcs) \cite{doi:10.1126/science.aaa9297,PhysRevX.5.031013,PhysRevApplied.18.054004,Huang2015,Yuan2018}. Interesting physical phenomena observed in these canonical Weyl semimetals include the chiral anomaly in charge transport, transport in the quantum limit under high magnetic fields, phase transitions between different topological states, strong intrinsic spin Hall effect, and symmetry-induced non-trivial spin-orbit torque \cite{yuan2020,PhysRevX.10.011050,Ramshaw2018,PhysRevApplied.18.054004,PhysRevLett.117.146403}. Most prior studies of this family of topological semimetals have used bulk single crystal samples; while these are of high structural quality, thin film samples are more desirable if these Weyl semimetals are to become useful for technological applications in microelectronics or optoelectronics. This provides a strong motivation for heterointegration of TX thin films with materials compatible with semiconductor processing; thin film growth would also enable the modulation of the band structure by controlling strain in the crystal \cite{Xu2015,PhysRevLett.116.096801}.

In this study, we develop the synthesis of NbAs thin films on a GaAs substrate using molecular beam epitaxy (MBE). We show that the growth of NbAs along different crystalline directions can be stabilized by judicious choice of the substrate direction. We then use x-ray diffraction to understand the macroscopic structure of the films, in addition to characterization using atomic force microscopy and  electrical transport measurements. Finally, we use high-angle annular dark-field scanning transmission electron microscopy (HAADF-STEM) to show that our films are textured with domains that show a clear crystal structure and are tens of nanometers in size.

%NbAs is a member of the TaAs family of Weyl semimetals 

%\section{\label{sec:level1}NbAs thin films grown on III-V substrates}

We first discuss the choice of substrate for the growth of NbAs thin films. NbAs is a transition metal monopnictide that has 12 pairs of Weyl nodes in its band structure \cite{Huang2015,Xu2015,doi:10.1126/science.aaa9297}. It is a member of the $\mathrm{I4_{1}md}$ space group and crystallizes in a body-centered tetragonal structure (Fig. 1(a)) with a lattice constant of $a = 0.345$ nm and $c=1.168$ nm \cite{Xu2015,doi:10.1126/science.aaa9297,PhysRevX.5.031013}. It is difficult to find an adequate substrate with a lattice constant in this range \cite{doi:10.1063/1.1368156bandgap}. Based on previous reports of MBE growth of this family of semimetals \cite{PhysRevApplied.18.054004,doi:10.1021/acs.cgd.2c00669,Bedoya-Pinto_ACSNano,https://doi.org/10.48550/arxiv.2303.05469}, we tried using different III-V semiconductors (GaAs, GaP and InAs) as our substrates. Among these, the best results were achieved in GaAs which had a lower lattice mismatch than InAs and, unlike GaP, had the right surface chemistry to promote NbAs growth. In this paper, we focus on studying NbAs films grown on GaAs; further information about characterization of NbAs on other substrates can be found in the supplemental material \cite{supp}.  

\begin{figure}
\includegraphics[width=130mm]{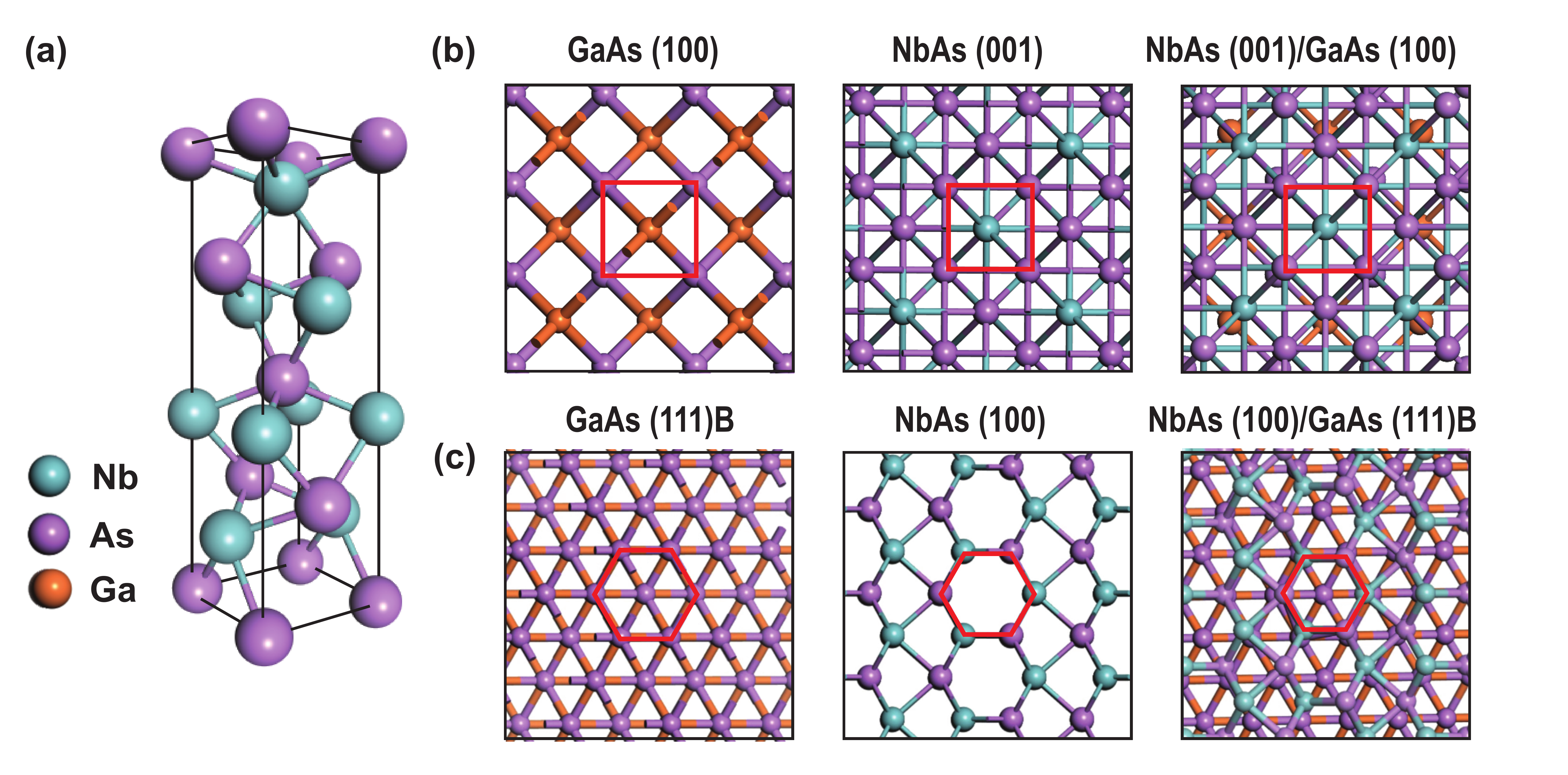}% Here is how to import EPS art
\caption{\label{fig:1} (a) Crystal structure of NbAs. (b) Top view of GaAs (100), NbAs (001) and NbAs (001) on top of GaAs (100). (c) Top view of GaAs (111)B, NbAs (100) and NbAs (100) on top of GaAs (111)B. We show the square and hexagonal symmetry and lattice mismatch between the films and the substrates.}
\end{figure}

An analysis of the crystal structure of NbAs suggests that the growth of NbAs (001) can be stabilized by using GaAs (100) as a substrate (Fig. 1(b)): here, the lattice mismatch is $\mathrm{16\%}$ along the $\langle 110 \rangle$ and $\mathrm{18\%}$ along the $\langle 100 \rangle$ GaAs directions. Although this is a very large lattice mismatch within the context of epitaxial growth, it is within range to allow NbAs to grow with a $\mathrm{45^\circ}$ rotation with respect to the $\langle 100 \rangle$ GaAs direction. NbAs (100) has a similar relationship with GaAs(111) as a substrate: Fig. 1(c) shows that if we superimpose both lattices, the NbAs atoms roughly align with the hexagonal structure of GaAs(111) with a $\mathrm{17\%}$ mismatch on each side.

%in the 100 and 111 directions 
%GaAs (100) and (111)B as our substrates. The lattice mismatch is quite significant between these materials. For TaAs (001) grown on GaAs (100) it is $\mathrm{16\%}$ along the $\langle 110 \rangle$ and $\mathrm{18\%}$  along the $\langle 010 \rangle$ GaAs directions. Nevertheless, we expect that GaAs will have the right surface chemistry to promote NbAs growth. We note that the MBE growth of similar Weyl semimetals (NbP and TaP) have been reported using MgO (001) substrates \cite{Bedoya-Pinto_ACSNano}. In this case a thin Nb buffer layer was needed to stabilize the film growth on a system with lattice mismatch as high as $9\%$.
%NbAs (001) film and the GaAs (100) substrate and the (b) NbAs (100) film and the GaAs (111)B substrate.
\begin{figure}
\includegraphics[width=80mm]{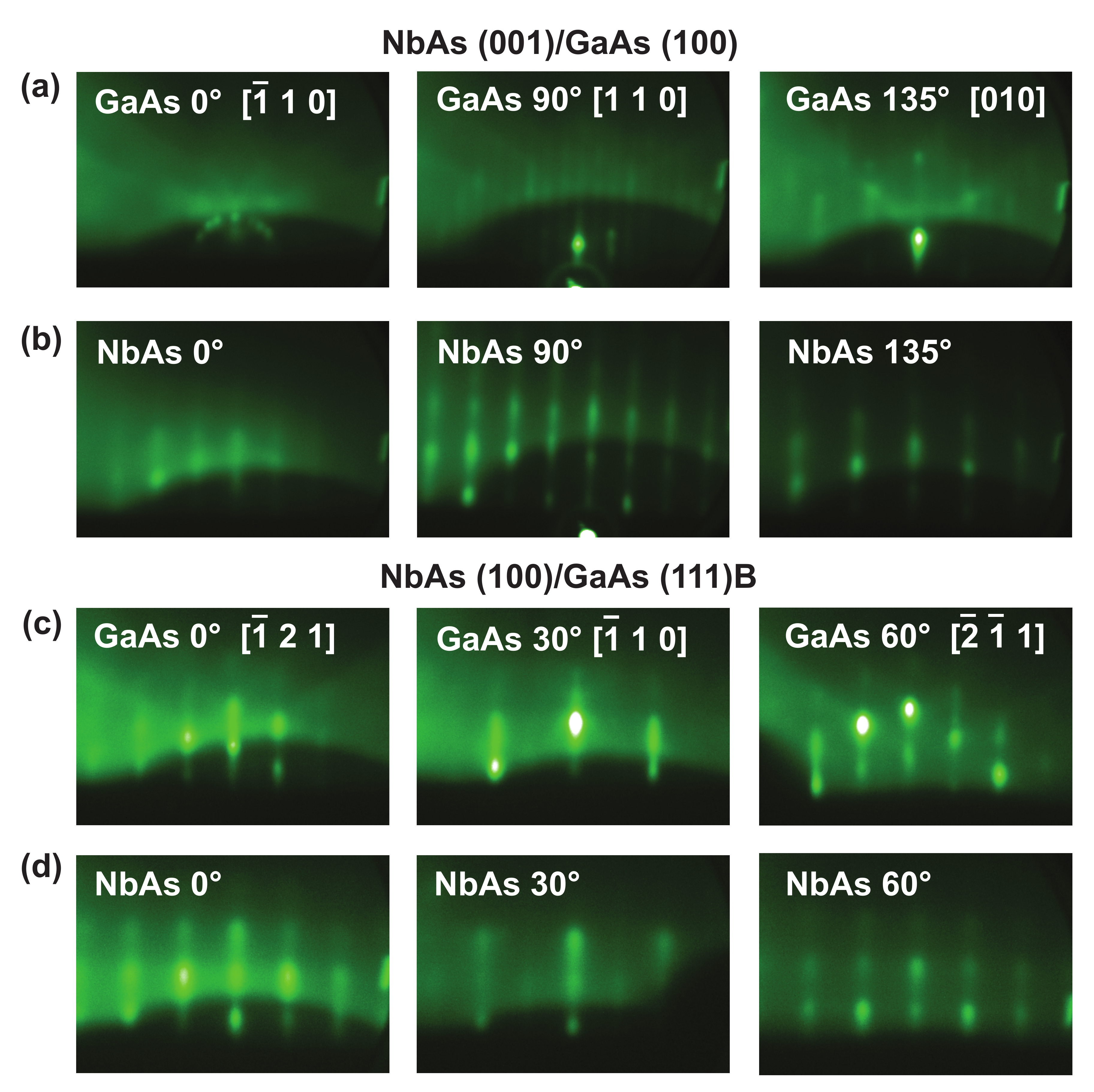}% Here is how to import EPS art
\caption{\label{fig:2} RHEED pattern of: (a) GaAs (100) substrate along the [$\mathrm{\bar{1}10}$], [110] and [010] directions. (b) NbAs (001) thin film along the same directions as (a), (c) GaAs (111)B substrate along the [$\mathrm{\bar{1}21}$], [$\mathrm{\bar{1}10}$] and [$\mathrm{\bar{2}\bar{1}1}$] directions, and (d) NbAs (100) thin film along the same directions as (c). }
\end{figure}

%\section{\label{sec:level1}Molecular beam epitaxy growth of NbAs thin films}

We carry out the synthesis of the NbAs films in a VEECO 930 MBE chamber while monitoring the growth using reflection high energy electron diffraction (RHEED) at 12 keV (Fig. 2). We desorb the native oxide on an epiready semi-insulating GaAs substrate and then grow 30 nm of GaAs at a thermocouple temperature of 720 $\mathrm{^\circ C}$ using Ga (5N) and As (5N) sources evaporated from standard effusion cells. The As:Ga beam quivalent pressure ratio is $\sim 14$ (as measured using an ion gauge). After this, we cool down the substrate to a thermocouple temperature of 400 $\mathrm{^\circ C}$ in the presence of As flux. At this point, we observed a RHEED pattern showing a $2 \times 4$ surface reconstruction for GaAs (100) (Fig. 2(a)) and a C6 symmetry for GaAs (111)B (Fig. 2(c)). We then increase the substrate temperature to 700-750 $\mathrm{^\circ C}$ (measured by a thermocouple in the substrate manipulator) and simultaneously deposit As and Nb (the latter from a SPECS EBE-4 4 pocket e-beam evaporator), obtaining the RHEED pattern shown in Figs. 2(b) and 2(d). In the case of NbAs grown on GaAs (100) (Fig. 2(b)), the RHEED pattern is different in the [110] and $\mathrm{[1\bar{1}0]}$ directions. This follows the C2 symmetry of the GaAs substrate in the $2 \times 4$ reconstruction. In the case of NbAs grown on GaAs (111)B (Fig. 2(c) and 2(d)), the NbAs RHEED pattern is the same in the GaAs $\mathrm{[1\bar{2}1]}$ and $\mathrm{[2\bar{1}\bar{1}]}$ crystal directions which are $\mathrm{60^\circ}$ apart. This indicates a C6 symmetry in the NbAs thin film due to twinning during growth. 
%For these reason, we follow a procedure similar to our previous report of TaAs MBE growth on a GaAs substrate \cite{yaneztaas}. 

%\section{Macroscopic X-ray characterization of the NbAs thin films}
% which is composed of a series of 360$^\circ$ $\phi$ scans at different $\chi$ angles. We aligned the film along the NbAs $\{112\}$ direction ($\chi=45.4^\circ$ and $\phi=46.2^\circ$) and proceed to scan around these peaks (Fig. 3(c) and 3(d)).
To characterize our films on a macroscopic scale, we use x-ray diffraction. First, we perform a coupled $\mathrm{2\theta-\omega}$ scan on NbAs grown on GaAs (100) (Fig. 3(a)) and detect the presence of diffraction peaks from NbAs (001). Second, we perform a similar scan on NbAs grown on GaAs (111)B (Fig. 3(b)) and detect the presence of NbAs (100). We emphasize that these different crystalline planes are not equivalent due to the tetragonal crystal structure of NbAs (Fig. 1). To quantify the degree of crystallinity in our films, we perform a rocking curve ($\omega$ scan) around the NbAs peaks and obtain a full width half maximum (FWHM) of 1.3$^\circ$ for NbAs (001) and 2.1$^\circ$ for NbAs (100). To understand the planar structure of our films, we measure a pole figure around the NbAs $\{112\}$ direction. We observe the presence of two different sets of peaks corresponding to GaAs $K\alpha$ and NbAs $K\beta$ radiation. We find that NbAs (001) stabilizes with a $45^\circ$ rotational offset with respect to the basal plane (Fig. 3(c)). As stated before, this is due to the lower lattice mismatch in the GaAs $\langle 110 \rangle$ direction. This agrees with previous reports of TaP and NbP grown on a substrate with cubic crystal structure and high lattice mismatch \cite{Bedoya-Pinto_ACSNano}. The difference in intensity between the peaks at $0^\circ$ and $90^\circ$ suggests that there is a preferred direction of alignment of the domains in the sample, probably due to the 2 $\times$ 4 GaAs surface reconstruction achieved before the growth. In Fig. 3(d), we see the presence of 3 dominant GaAs peaks with the expected C3 symmetry of GaAs in the (111) direction. We also see the presence of 6 NbAs peaks that are 60$^\circ$ apart. This indicates the presence of domains (twins) that are rotated by 60$^\circ$ with respect to each other. Finally, to understand the structure of the surface of the films, we use atomic force microscopy (AFM) imaging on these samples. In both surfaces (Figs. 3(e) and 3(f)) we see domains that are tens of nanometers in size. The main difference between these different surfaces is that in the (001) plane we see the presence of a rougher surface with domains that show the tetragonal structure of NbAs while in the (100) direction, we have a smoother surface with a higher degree of disorder. This qualitatively agrees with the larger FWHM obtained in the rocking curve in NbAs (100).

\begin{figure}

\includegraphics[width=120mm]{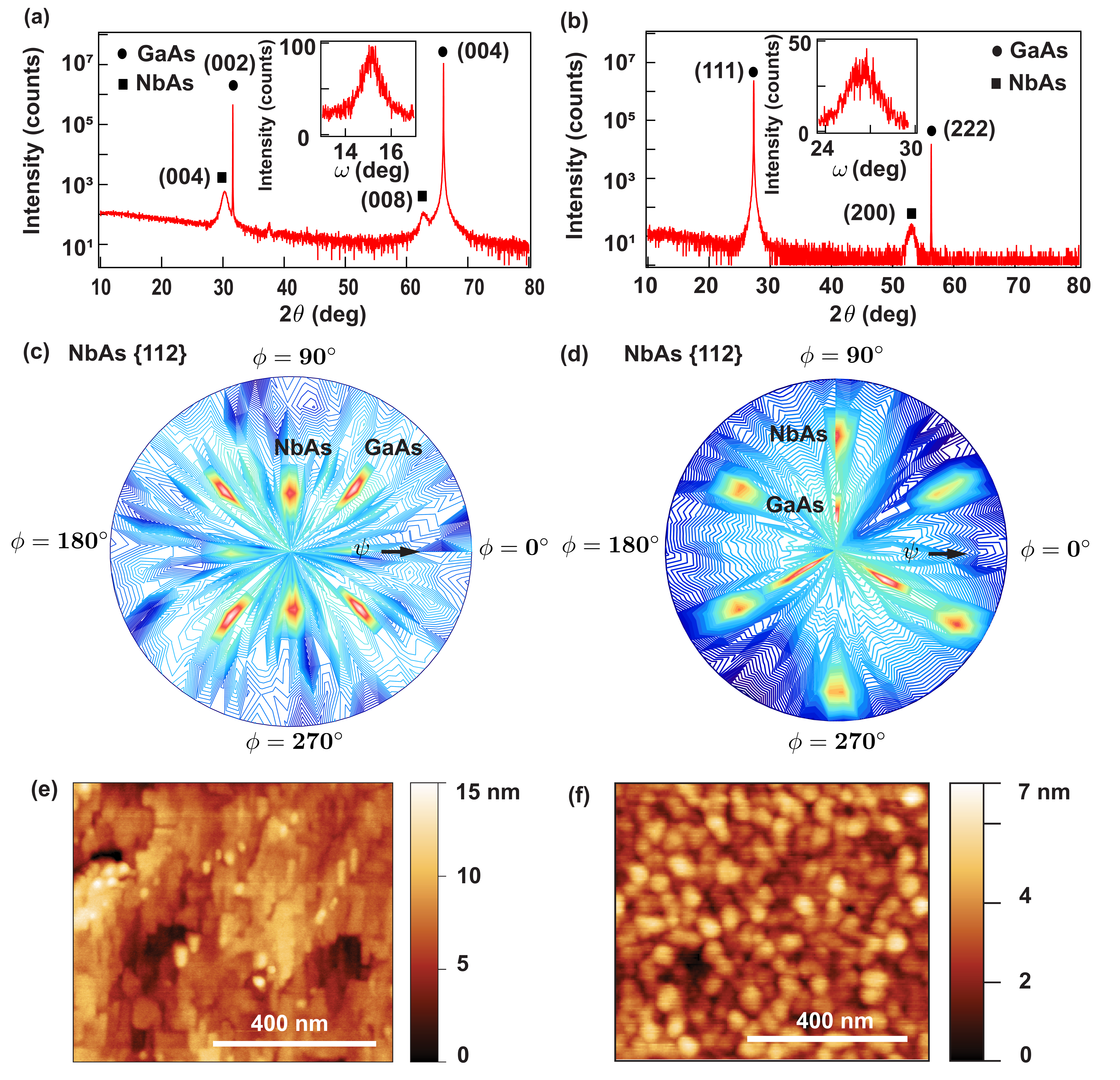}% Here is how to import EPS art
\caption{\label{fig:3} X-ray diffraction $\mathrm{2\theta-\omega}$ scan of NbAs grown on (a) GaAs (100) and (b) GaAs (111)B. Pole figures around the NbAs $\{112\}$ direction for samples grown on (c) GaAs (100) and (d) GaAs (111)B. Atomic force microscopy image of (e) NbAs (001) and (f) NbAs (100) showing a root mean square (RMS) surface roughness of 1.4 nm and 0.9 nm respectively. }
\end{figure}

To understand the microscopic structure of NbAs in the thin film regime, we use HAADF-STEM imaging and energy-dispersive X-ray spectroscopy (STEM-EDX). The results shown in the present manuscript correspond to NbAs (001) grown on GaAs (100). Similar measurements are obtained for NbAs (100)/GaAs (111)B.  Figure 4 (a), shows a cross-section HAADF-STEM image of NbAs on GaAs. The NbAs layer is polycrystalline with grain sizes in the range of tens of nanometers, in agreement with the XRD results. The interface between the NbAs and GaAs is diffuse and shows interdiffusion, likely due to the high temperatures required to grow NbAs and the large lattice mismatch with GaAs. The relative concentration of Nb:As and Ga:As is 1:1 in both layers, which confirms the absence of other phases (Fig. 4 (b) and (c)). There is a self-limited oxide layer that covers the top $\sim$3 nm of the NbAs film. This is consistent with our previous studies of other topological semimetals such as TaAs and $\mathrm{Cd_3As_2}$ \cite{PhysRevApplied.18.054004,PhysRevApplied.16.054031yanez}. 

\begin{figure}
\includegraphics[width=130mm]{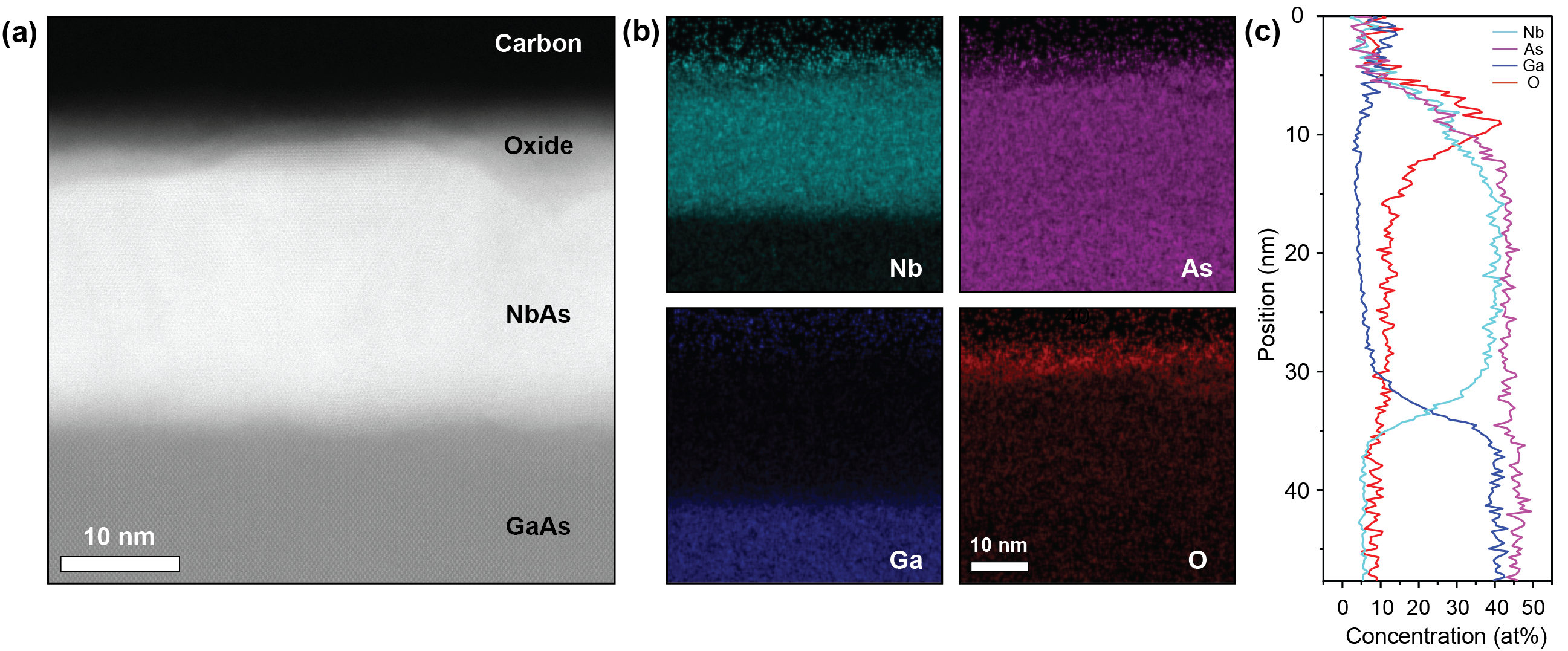}% Here is how to import EPS art
\caption{\label{fig:5} (a) HAADF-TEM image of the NbAs (001) thin film on a GaAs substrate. (b) STEM-EDX elemental maps of Nb, As, Ga and O in the film. (c) Concentration of the elements in (a) across the NbAs/GaAs interface.}
\end{figure}

The polycrystalline NbAs layer has domains oriented along the major crystal directions, in addition to the presence of some grain boundaries and twins (Fig. 5(a)). Two such examples are shown in Fig. 5 (b) which are commonly observed in the thin films grown on GaAs.  Comparison of the fast Fourier transforms (FFTs) from the grains to the simulated FFTs confirms the two grains to be along the (001) and (100) directions respectively (Fig. 5 (c) and (d)).  We note that this is a small subset of several different domains that were identified in NbAs that had more complex orientations that were harder to identify. We believe that more complex microscopy techniques that go beyond the scope of this work (i.e. in plane imaging or statistical analysis of domain orientation) are needed to fully understand the structure of the domains in NbAs thin films.

\begin{figure}
\includegraphics[width=130mm]{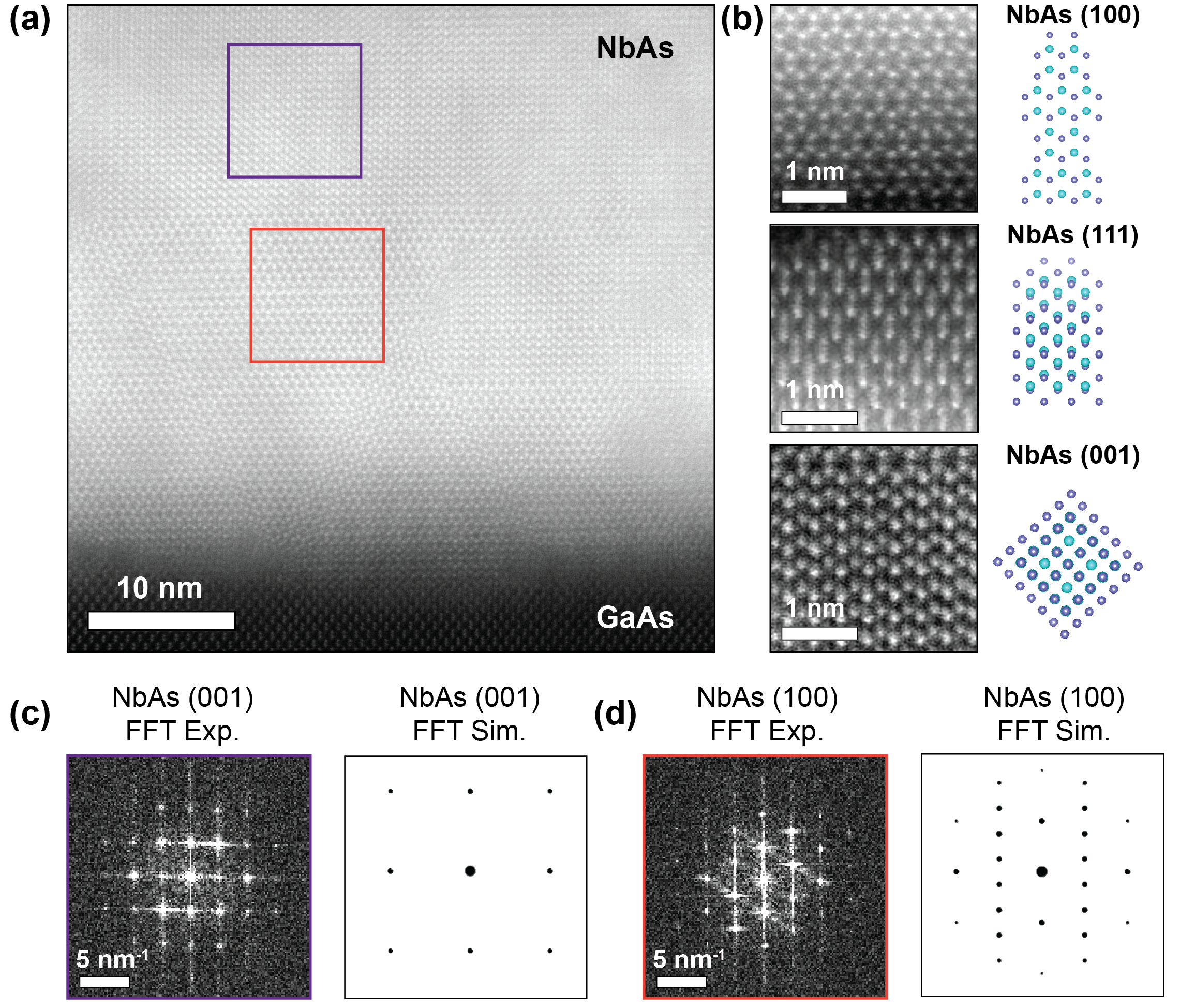}% Here is how to import EPS art
\caption{\label{fig:6} (a) HAADF-STEM image of NbAs showing polycristalline structure with multiple  grains along different crystal orientations. (b) Common NbAs orientations observed in grains along with the theoretical model. (c) and (d) FFT of the NbAs grains highlighted by the purple and orange boxes respectively along with the simulated FFTs for the  (001) and (100) orientations.}
\end{figure}

%\section{Electrical transport in NbAs thin films}

In order to characterize the electrical properties of the NbAs thin films, we fabricated Hall bars using standard photolithography and Ar plasma etching. The dimensions of the films (length, width and thickness) are $1000$~$\mu$m x $500$~$\mu$m x 23 nm and $40$~$\mu$m x $10$~$\mu$m x 10 nm respectively. Both films show similar qualitative and quantitative behavior. The resistivity ($\rho$) vs. temperature ($T$) in both samples shows activated behavior (Fig. 6(a) and (b)), possibly arising from defects, with $\rho\approx450~\mu\Omega$ cm. Hall measurements performed at $T=2$~K show electrons to be the dominant carrier in both samples (Fig. 6(c) and (d)), with a carrier density ($n$) $6.5\times10^{22}$~cm$^{-3}$ for NbAs (001) and $5\times10^{21}$~cm$^{-3}$ in NbAs (100). The Hall mobility is $0.2$ and $2$~cm$^{-2}$Vs respectively, while $k_fl=2$ and $5$ respectively ($k_f$ is the fermi momentum and $l$ is the mean free path), implying highly disordered films. Longitudinal resistance measurements at $T=2$~K show positive magneto-resistance in both samples (Fig. 6(c) and (d)). 
Although $k_fl$ is close to the limit for diffusive transport conditions, as set by the Ioffe-Regel limit ($k_fl \approx1$), we estimated the phase breaking length ($l_\phi$) in both films using the Hikami-Larkin-Nagaoka expression~\cite{10.1143/PTP.63.707} for weak-antilocalization for high spin-orbit coupled systems:
\begin{equation}
\triangle\sigma=\alpha\frac{e^{2}}{\pi h}\left[\psi\left(\frac{1}{2}+\frac{B_{\phi}}{B}\right)-\ln\left(\frac{B_{\phi}}{B}\right)\right]\label{eq:HLN}
\end{equation}
Here, $B_{\phi}$ is the phase coherence field, and $\alpha$ is a fitting parameter. We estimate the phase breaking length ($l_\phi=\sqrt{\frac{\hbar}{4eB_\phi}}$) in both films, to be $50-80$~nm.
 
%signatures consistent with weak-antilocalization, as expected for a material with strong spin-orbit coupling  %\cite{10.1143/PTP.63.707}.
%Even though disorder is strong in our samples and they do not satisfy the Ioffe-Regel criterion (the product of the %Fermi momemtum and the mean free path is $k_Fl=2$ and $5$ respectively), we tried to 

%We note that significant differences in geometrical aspects between these two samples makes it difficult to do a fair comparison among them. In future work, we expect to perform detailed electrical characterization and more fully understand the electronic transport properties of NbAs. 

 %\textcolor{blue}{The Hall mobility in both films are $0.2$ and $2$~cm$^{-2}$Vs respectively, while $k_fl=2$ and $5$ respectively.} Longitudinal magnetoresistance measurements at $T=2$~K show weak-antilocalization as expected for high spin-orbit coupled systems (fig. 4(c) and (d)) [CITE]. Using Hikami-Larkin-Nagaoka expression.}
%\begin{equation}
%\triangle\sigma=\alpha\frac{e^{2}}{\pi h}\left[\psi\left(\frac{1}{2}+\frac{B_{\phi}}{B}\right)-\ln\left(\frac{B_{\phi}}{B}\right)\right]\label{eq:HLN}
%\end{equation}
%\textcolor{red}{Here, $B_{\phi}$ is the phase coherence field, and $\alpha$ is a fitting parameter., we estimate the phase breaking length ($l_\phi=\sqrt{\frac{\hbar}{4eB_\phi}}$) in both films, to be $50-80$~nm, consistent with the textured nature of the films. }

\begin{figure}
\includegraphics[width=120mm]{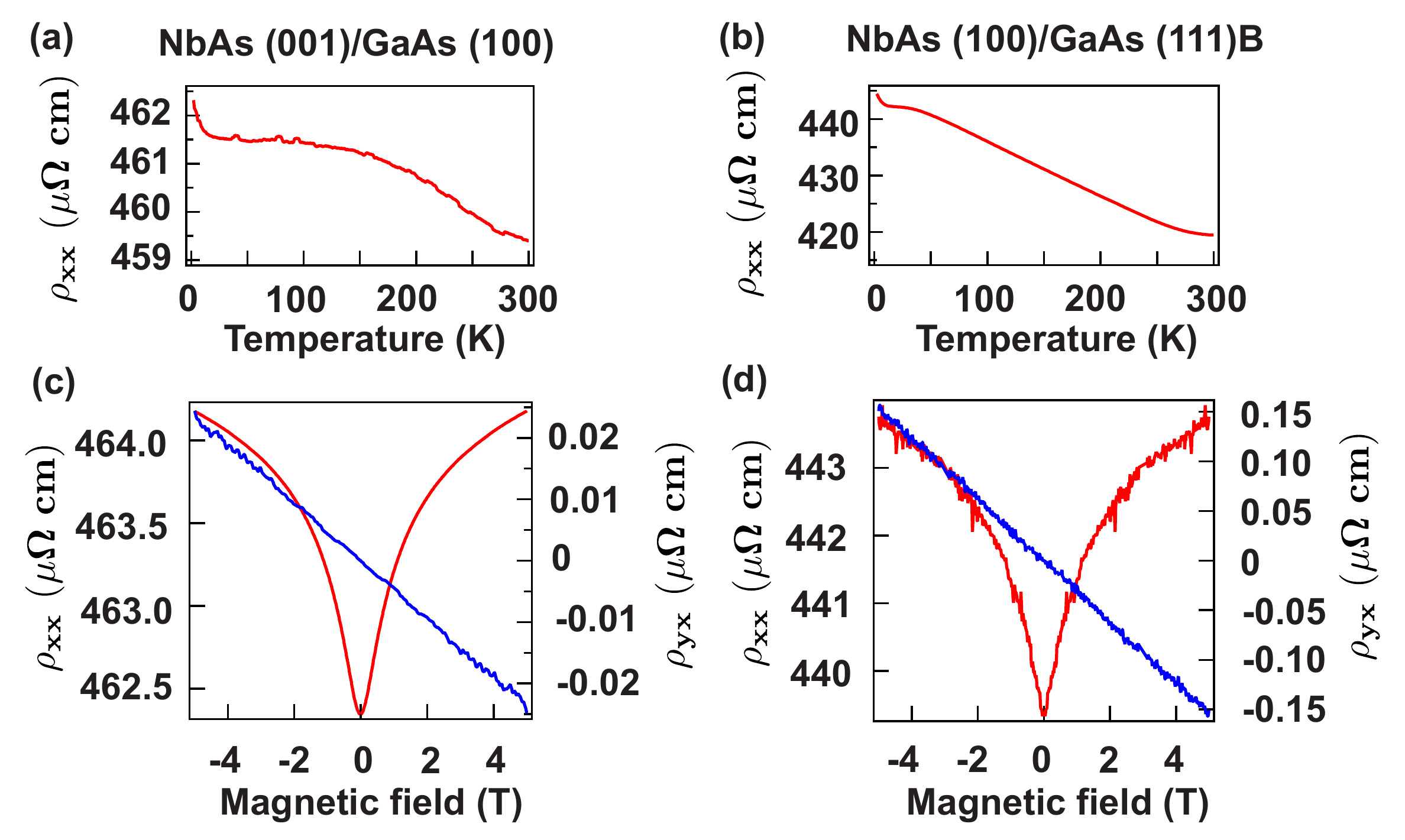}% Here is how to import EPS art
\caption{\label{fig:4} (a) and (b) Longitudinal resistivity ($\rho_{xx}$) as a function of temperature in 23 nm thick NbAs (001) and 10 nm thick NbAs (100). (c) and (d) Longitudinal resistivity ($\rho_{xx}$) and Hall resistivity ($\rho_{yx}$) as a function of magnetic field in NbAs (001) and (100). }
\end{figure}

%\section{Conclusions}

In summary, we report on the synthesis by MBE of Weyl semimetal NbAs thin films on GaAs substrates, a semiconductor of relavence for optoelectronics. We show that different NbAs crystalline orientations can be stabilized by choosing the appropriate substrate orientation. The large lattice mismatch between GaAs and NbAs, coupled with the high growth temperature needed to achieve NbAs growth, induces diffusion at the GaAs/NbAs interface. This hinders the crystallinity of our samples and produces textured films with domains that are tens of nanometers in size. These domains nonetheless have a well defined crystal structure that is oriented in different crystalline directions. Despite the large lattice mismatch, the surface chemistry of GaAs seems to be a decisive factor for effectively nucleating NbAs growth. There is some apparent inconsistency between the textured nature of our films shown in STEM and the clear peaks corresponding to a single crystalline orientation shown in the X-ray diffraction data. We reconcile both measurements by proposing the existence of a preferred crystalline orientation of these domains that can only be seen on a macroscopic scale. More detailed microscopy studies that include statistical analysis and in plane imaging, are required to confirm this hypothesis. We observe that the films naturally oxidize when taken out of vacuum. This effect makes it difficult to access the electronic states of NbAs using surface sensitive techniques like angle resolved photoemission spectroscopy or scanning tunneling spectroscopy that involve the out-of-vacuum transfer of samples between different vacuum chambers. This suggests caution when trying to probe topological surface states in this family of Weyl semimetals in samples with an oxidized surface. Finally, we expect that further experimental improvements will allow us to increase the degree of crystallinity in NbAs films so that we are able to explore a broader range of physics in this family of well-established Weyl semimetals.

%It has been shown that the presence of this oxidized layer can have dramatic effects on the properties of similar films in this family of Weyl semimetas \cite{yaneztaas}. For this reason, we beg caution in further studies that aim to probe the topological surface states of this family of Weyl semimetals in samples with an oxidized surface. 

%even though similar analysis in NbAs goes beyond the scope of this paper. We point to the fact that oxidation also happens in the surface of NbAs and it should be suppressed in further studies that aim to probe the topological surface states of this family of Weyl semimetals.

%This has been confirmed with thorough X-ray diffraction together with transmission electron microscopy 

%In spite of the large lattice mismatch, the surface chemistry of GaAs seems to be a decisive factor to stimulate NbAs growth. This mismatch together with the high growth tempeature that induces diffusion at the GaAs/NbAs interface causes the emergence of domains with a well defined crystal structure that are   tenths of nanometers in size. This has been confirmed with thorough X-ray diffraction together with transmission electron microscopy characterization. 

%show that the films are textured with tenths of nanometer size .   

%\begin{acknowledgments}
The principal support for this project was provided by SMART, one of seven centers of nCORE, a Semiconductor Research Corporation program, sponsored by the National Institute of Standards and Technology (NIST). This supported the synthesis and standard characterization of NbAs films (WY, YH, NS) and their characterization using STEM (SG, AM). Additional support for materials synthesis and characterization was provided by the Penn State Two-Dimensional Crystal Consortium-Materials Innovation Platform (2DCC-MIP) under NSF Grant No. DMR-2039351 (SI, ES, AR, NS). Parts of this work were carried out in the Characterization Facility, University of Minnesota, which receives partial support from the NSF through the MRSEC (Award Number DMR-2011401) and the NNCI (Award Number ECCS-2025124) programs. SG acknowledges support from a Doctoral Dissertation Fellowship received from the Graduate School at the University of Minnesota.  This work utilizes low-temperature transport facilities provided by the Penn State Materials Research Science and Engineering Center under award NSF-DMR 2011839.
%\end{acknowledgments}

%\section*{Data Availability Statement}

%The data that support the findings of this study are available
%from the corresponding authors upon reasonable request.

%\nocite{*}
%\bibliography{aipsamp}% Produces the bibliography via BibTeX.

%bibliography

\newpage

\providecommand{\noopsort}[1]{}\providecommand{\singleletter}[1]{#1}%

 \end{document}